# Dynamic Skyrmion-Mediated Switching of Perpendicular MTJs: Scaling to 20 nm with Thermal Noise

Md Mahadi Rajib, Walid Al Misba, Dhritiman Bhattacharya, Student Member, IEEE, Felipe Garcia-Sanchez, Member, IEEE and Jayasimha Atulasimha, Senior Member, IEEE.

*Abstract*—One method of creating and annihilating skyrmions in confined geometries is to use Voltage-Controlled Magnetic Anisotropy (VCMA) [1, 2, 3]. Previous study shows that robust voltage controlled ferromagnetic reversal from "up" to "down" state in the soft layer of a perpendicular Magnetic Tunnel Junction (p-MTJ) can be achieved by creating and subsequently annihilating an intermediate skyrmion state [4] in the presence of room temperature thermal noise and anisotropy variation across grains [4]. However, when scaling to 20 nm, thermal noise can annihilate the skyrmions, for example, by randomly moving the core towards the boundary of the nanostructure. In this work, we study three p-MTJs of different dimensions, particularly lateral dimensions of 100nm, 50nm and 20 nm and investigate the change in switching behavior as the dimension is decreased. Particularly, our focus is to investigate to what extent skyrmion mediated switching scheme can be employed below ~50nm lateral dimensions in the presence of thermal perturbation.

## I. INTRODUCTION

Typically, nanomagnetic memory elements are based on perpendicular Magnetic Tunnel Junctions (p-MTJ) as shown in Fig. 1 (a). The magnetoresistance of an MTJ can be changed by switching the magnetization direction of the free layer between two stable orientations (i.e. writing a memory bit), which in turn allows reading out of the memory bit [5-8]. Various current induced [9-13] and electric field-controlled [14-22] mechanisms can be used for switching the free layer of an MTJ. Spin transfer torque (STT) [9, 10], one of the most commonly employed current controlled switching strategies, causes a large energy dissipation (~100fJ/bit). For comparison, energy dissipated in CMOS devices is approximately 100 aJ/bit [23], which is nearly three orders of magnitude lesser than energy dissipated in STT based switching. On the other hand, electric field-induced magnetization reversal is highly energy efficient. In particular, Voltage Controlled Magnetic Anisotropy (VCMA) based switching dissipates only ~1fJ/bit [18, 24]. In this method, a voltage pulse alters the perpendicular magnetic anisotropy originating at the interface of ferromagnet/oxide [25]. Utilizing an appropriate combination of in-plane bias magnetic field and voltage pulse duration, complete ferromagnetic reversal can be achieved [15]. However, VCMA induced switching is precessional in nature and highly susceptible to disorders in the presence of room-temperature thermal noise [4].

Recently, the possibility of VCMA induced magnetization reversal between two ferromagnetic states without any external bias magnetic field has been reported [4]. This is achieved by applying a voltage pulse that reduces PMA. Thus, starting from one of the ferromagnetic states (up/down), an intermediate skyrmion can be created. The voltage pulse is withdrawn coincidentally with the inbreathing motion of the skyrmion. Due to this, when the initial PMA value is restored, the intermediate skyrmion annihilates to the other [4] ferromagnetic state (down/up). The switching mechanism is shown in Fig. 1(b). Skyrmions are usually controlled by electrical current in racetrack devices which was proven to be advantageous due to its potential in overcoming edge roughness related pinning occurring in Domain Wall (DW) based racetrack devices [26-28]. However, the skyrmion-mediated switching strategy explored in this paper uses a confined geometry and thus has potential for higher density memory cells compatible with crossbar architecture. This mechanism has been shown to be robust to thermal noise, disorder and

Submitted on March xx, 2020. M. M. R., D.B. and J.A. are supported by NSF SHF Small Grant #1909030.

M. M. Rajib, W. Misba and D. Bhattacharya are with the Department of Mechanical and Nuclear Engineering, Virginia Commonwealth University, Richmond, VA 23284 USA (e-mail: rajibmm@mymail.vcu.edu, misbawa@mymail.vcu.edu and bhattacharyad@vcu.edu).

F. Garcia-Sanchez is affiliated with Departamento de Física Aplicada, University of Salamanca, Pza de la Merced s/n, 37008 Salamanca, Spain (f.garcia@inrim.it)

J. Atulasimha is with the Department of Mechanical and Nuclear Engineering and the Department of Electrical and Computer Engineering, Virginia Commonwealth University, Richmond, VA23284 USA (e-mail: jatulasimha@vcu.edu)

perturbative spin currents for a nanomagnet of lateral dimension 100nm. However, to be competitive with the current miniaturization trend in Spin Transfer Torque Random Access Memory (STTRAM), further downscaling is required. In this work, we theoretically investigate the feasibility of downscaling the skyrmion mediated switching scheme by studying three MTJs of different sizes, 100nm, 50 nm and 20 nm respectively. We observe that with the reduction in lateral dimension with concomitant scaling of the thickness, high perpendicular magnetic anisotropy, high DMI and a larger VCMA coefficient are required for successful operation of the device. The breathing frequency of skyrmions in the downscaled nanomagnets is high leading to switching frequency ~1 GHz, ~10 GHz, ~100 GHz for 100 nm, 50 nm and 20 nm respectively. Thus, as one scales to lateral dimension less than 50 nm, there is likely to be a very narrow range of pulse width where the switching is robust in the presence of thermal noise. There are many more interesting insights such as material parameters required, switching error, physical mechanism for this error, etc., that are studied in this paper. Section II of this paper discusses the method followed for numerical simulations in this study. Section III describes the findings from this study and Section IV makes concluding remarks based on the numerical simulations on potential scaling in skyrmion mediated switching of p-MTJs.

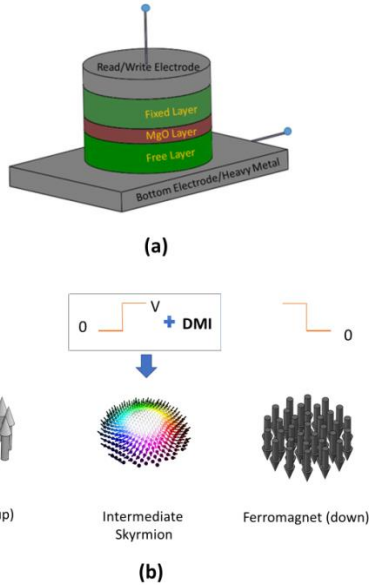

Fig. 1: (a) Schematic presentation of a Magnetic Tunnel Junction and (b) Skyrmion mediated switching scheme

## II. METHOD

We simulate the magnetization dynamics of the free layer of the MTJs, which were discretized into 64×64×1, 32×32×1 and 16×16×1 cells for 100nm, 50nm and 20nm MTJs respectively to keep the lateral (in-plane) discretization similar to the extent possible. The magnetization dynamics is simulated by solving the Landau-Lifshitz-Gilbert (LLG) equation via the micromagnetic simulation software Mumax3 [29]:

$$\frac{\partial \vec{m}}{\delta t} = \left(\frac{-\gamma}{1+\alpha^2}\right)\left[\vec{m} \times \vec{B}_{eff} + \alpha\{\vec{m} \times (\vec{m} \times \vec{B}_{eff})\}\right] \quad (1)$$

In eqn (1), $\gamma$ is the gyromagnetic ratio, $\alpha$ is the Gilbert damping coefficient. $\vec{m}$ is the normalized magnetization vector, found by normalizing the magnetization vector ($\vec{M}$) with respect to saturation magnetization ($M_s$) and $\vec{B}_{eff}$ is the effective magnetic field with the following components [29]:

$$\vec{B}_{eff} = \vec{B}_{demag} + \vec{B}_{exchange} + \vec{B}_{dm} + \vec{B}_{anis} + \vec{B}_{thermal} \quad (2)$$

$\vec{B}_{demag}$ is the effective field due to demagnetization energy, $\vec{B}_{exchange}$ is the effective field due to the Heisenberg exchange interaction and $\vec{B}_{dm}$ is the effective field due to Dzyaloshinskii-Moriya interaction.

The effective field due to the perpendicular anisotropy, $\vec{B}_{anis}$ is given by the following equation:

$$\vec{B}_{anis} = \frac{2K_{u1}}{M_s}(\vec{u}.\vec{m})\vec{u} + \frac{4K_{u2}}{M_s}(\vec{u}.\vec{m})^3\vec{u} \quad (3)$$

$K_{u1}$ and $K_{u2}$ are the first and second order uniaxial anisotropy constants and $\vec{u}$ a unit vector in the anisotropy direction. VCMA induced PMA modulation is employed by modulating $K_{u1}$ while keeping $K_{u2} = 0$

Thermal field contribution to the effective field is calculated by:

$$\vec{B}_{thermal} = \vec{\eta}(step)\sqrt{\frac{2\mu_0 \alpha k_B T}{M_s \gamma \Delta V \Delta t}} \quad (4)$$

$k_B$ the Boltzman constant, T the temperature, $\Delta V$ the cell volume, $\Delta t$ is time step which is considered to be 10fs for all of our simulation cases (reason explained in the appendix) and $\vec{\eta}(step)$ a random vector from a standard normal distribution whose value is changed after every time step.

## III. RESULTS AND DISCUSSION

Our proposed structure is presented in Fig. 1(a). The free layers were chosen to be nanodisks of diameter 100 nm, 50 nm and 20 nm and thickness of

1.2 nm, 0.6 nm and 0.24 nm respectively. The thickness to diameter ratio (the aspect ratio) was set to 0.012 for all three MTJs in order to maintain constant demagnetization factor. We considered exchange stiffness A=25 pJ/m, saturation magnetization $M_s=1.3\times10^6$ A/m, Gilbert damping coefficient α=0.01. For successful realization of our switching mechanism, we require an initial ferromagnetic state. Next, when the PMA is reduced (voltage applied), we need to form an intermediate skyrmion state to provide a robust pathway for magnetization reversal to the opposite ferromagnetic state when the PMA is restored (voltage withdrawn). We varied the perpendicular anisotropy ($K_{u1}$) and the DMI parameter (D) to fulfill these requirements. The rationale behind choosing different anisotropy and DMI values is discussed below.

Firstly, to form the intermediate skyrmion state, we need a sizeable DMI. Moreover, reduction of lateral dimension demands higher DMI to form an intermediate skyrmion state. However, to maintain an initial ferromagnetic state, the DMI value must be less than the critical value ($D_{crit}=\frac{4}{\pi}\sqrt{AK_{EFF}}$, where $K_{EFF}= K_{u1}-\frac{1}{2}\mu_0 M_s^2$) at the corresponding initial PMA values. Here, $K_{EFF}$ is the effective perpendicular anisotropy and the $D_{crit}$ (critical DMI) value signifies a threshold beyond which a metastable skyrmion can prevail in the system.

Energy barrier between two ferromagnetic states reduces due to the presence of DMI [30]. To incorporate this reduction, we have estimated the energy barrier with $K_{eff}$ from the phenomenological equation $K_{eff}= (K_{u1}-\frac{1}{2}\mu_0 M_s^2-\frac{D^2\pi^2}{16A})$. The perpendicular anisotropy is determined from the thermal stability factor $K_{eff}V/K_BT$, which was considered to be 50 for all three MTJs. Therefore, with reduction of lateral dimension, required DMI increases and consequently the uniaxial anisotropy $K_{u1}$ needs to be increased to ensure a thermal stability factor of 50.

The parameters are listed in Table I (A=25 pJ/m, $M_s=1.3\times10^6$ A/m, α=0.01 for all cases):

TABLE I

| MTJs | $K_{u1}$ (MJ/$m^3$) | $D_{crit}$ (mJ/$m^2$) | D (mJ/$m^2$) |
|---|---|---|---|
| 100nm | 1.11 | 1.39 | 1.2 |
| 50nm | 1.63 | 4.80 | 4.5 |
| 20nm | 7.98 | 16.74 | 13.0 |

We note that the feasibility of the material parameters such as PMA ($K_{u1}$) and DMI coefficient (D) in Table I (and also the VCMA coefficients mentioned on Table II later) are discussed in the conclusion section.

These parameters along with the switch error analysis form the basis for determining the extent to which this switching scheme can be downscaled.

### A. Non-thermal

First, we study the behavior of the nanomagnets by reducing PMA through VCMA without any thermal perturbation. At reduced PMA, the competing PMA, DMI and demagnetization forms a skyrmion. Once the skyrmion is formed, it starts to oscillate between two stable states: skyrmionic state and quasi-ferromagnetic state. The oscillation is observed in the form of topological charge evolution with time (Fig. 2) where the crests indicate a skyrmionic state and troughs indicate a quasi-ferromagnetic state. With the reduction of lateral dimension, the DMI required to form an intermediate skyrmion state increases. Thus, it forces the magnetization to oscillate at a higher frequency. This reduces the pulse width and the rise time of the voltage pulse applied that can produce favorable switching in the smaller nanomagnets. VCMA coefficients, fall and rise time, reduced PMA, oscillation frequency and pulse-width range are listed in Table II considering application of voltage pulse (ΔV) of 2.0 V across 1 nm thick MgO layer for all three MTJs.

TABLE II (Columns 2-4 are parameters chosen and column 5, 6 are inferred from simulations)

| MTJs | VCMA Coeff. (fJ/Vm) | Fall and rise time (ps) | $K_{u1}$ (MJ/$m^3$) | *Pulse-width range (ps)* | *Osc. Freq. (GHz)* |
|---|---|---|---|---|---|
| 100nm | 100 | 100 | 0.94 | *500-900* | *~1* |
| 50nm | 215 | 10 | 0.90 | *40-90* | *~8* |
| 20nm | 825 | 1 | 1.10 | *5-14* | *~60* |

By withdrawing the voltage pulse and restoring the PMA to the initial values at each peak point, switching is observed. Fig. 2 shows that the nanomagnets switch for the first 5, 10 and 4 cycles for 100nm, 50 nm and 20 nm lateral dimension (diameter) nanomagnets respectively. Beyond these cycles (for respective nanomagnet sizes), the topological number falls below the critical value and therefore an intermediate skyrmion state is not available for successful switching. The "switch" and "no switch" regions along with the comparison of topological evolution for three nanomagnets are shown in Fig. 2. By the "switch" region we mean that when the voltage pulse is withdrawn (PMA restored) roughly at the maxima of the topological charge in Fig. 2, the magnetization of the nanomagnet relaxes to a ferromagnetic state that is opposite (reversed) from the original state. In other words, the switching or reversal is successful. In the "no switch" region, even if the voltage pulse is withdrawn (PMA restored) roughly at the maxima of topological charge the magnetization relaxes to the original ferromagnetic state and thus does not switch successfully.

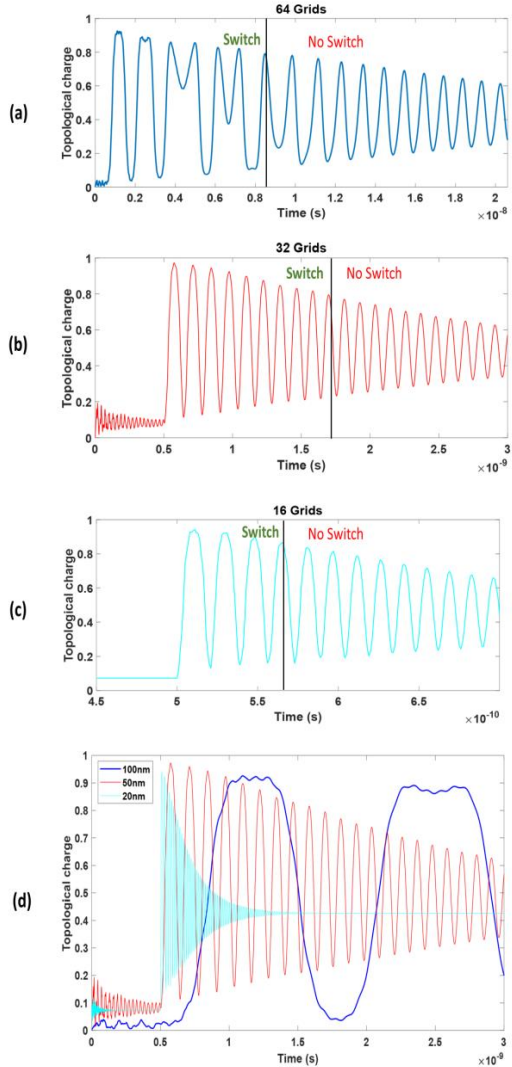

Fig. 2: "Switch" and "no switch" region without thermal perturbation for (a) 100nm (b) 50 nm (c) 20 nm and (d) Comparison of topological evolution of three nanomagnets using same time scale for the nanomagnets of three different sizes.

### B. Thermal

In the presence of room temperature thermal noise, the first breathing period of skyrmion is ideal for the switching to minimize the possibility of intermediate skyrmion annihilation due to thermal noise. The switching percentage for 100 nm nanomagnet with pulse width range of 400-700 ps is studied and shows thermally robust switching in the 450-650 ps range (Fig. 3a). For 50 nm, 97.5% switching can be attained for pulse width of 50 ps (Fig. 3b), which could be improved by optimizing the pulse shape and choice of material parameters. For 20 nm, switching percentage is studied for pulse width range of 5-14 ps. Within this range, the maximum switching percentage observed is only ~34% (this means of 100 simulations to toggle/reverse the magnetization only 34 switched successfully) for PW of 8 ps (Fig. 3c).

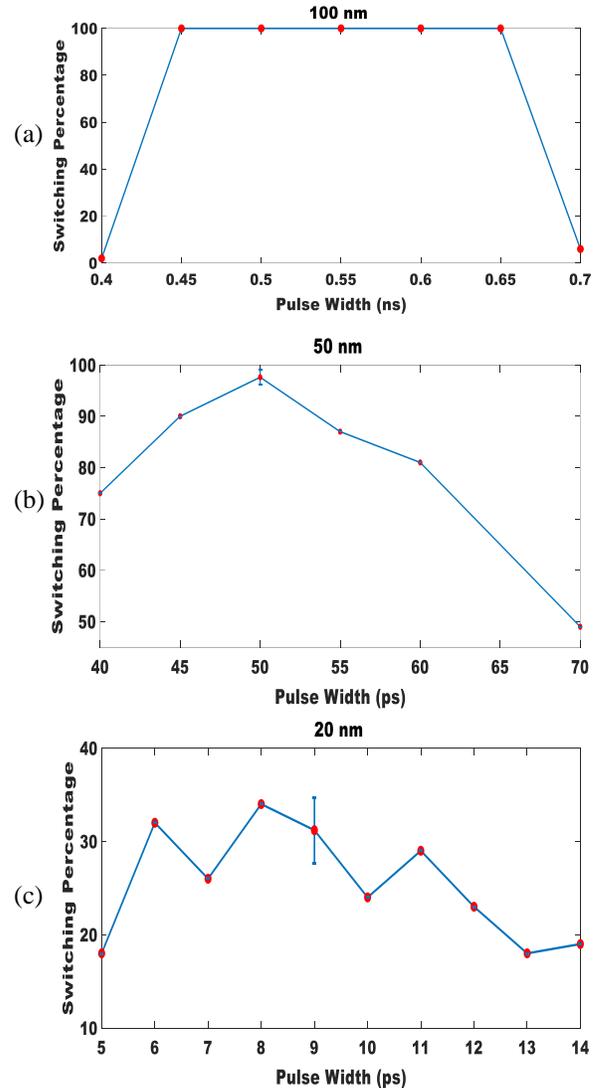

Fig. 3: Switching probability vs pulse width of (a) 100nm (b) 50nm and (c) 20 nm nanomagnets.

[For studying the switching percentage for different pulse widths, the simulations were run for 100 times for most points. For just one point in each sub-figure (corresponding to a specific pulse width for each nanomagnet of a different lateral dimension), an error bar indicates the simulations were run 1000 times. The points marked as 100% switch for the 100 nm nanomagnet implied no error for 100 simulations (or 1000 for 500 ps pulse width) and no error is bar is used as this was true for repeated batches of simulations. However, for the 50 nm and 20 nm error bars are used to denote standard deviation estimated from 10 batches of 100 simulations at pulse width of 50 ps and 9 ps respectively.]

We present the magnetization dynamics of the 20 nm nanomagnets in detail to explain why the switching error is very high in the presence of thermal noise. We first present the magnetization states visited during successful reversal of 20 nm nanomagnet without thermal noise in Fig. 4a. The initial state (0 ps) is a quasi-ferromagnetic state and spins are tilted slightly inward at the disk edge. This state is obtained starting with an upward (+z) pointing ferromagnetic state (initial condition) from which the simulation is run for 100 ps to allow it to relax to this equilibrium state. As the PMA is reduced, creation of a skyrmion can be observed (6ps,9ps) and with the withdrawal of voltage pulse, the magnetization completely switches to downward (-z) oriented ferromagnetic state (18 ps). The same qualitative behavior is also observed with thermal noise (Fig. 4 b): initial ferromagnetic upward orientation (0ps), formation of intermediate skyrmion (6ps, 9ps) and switched downward ferromagnetic state (18ps).

We next report micromagnetic states in the magnetization dynamics of the 20 nm nanomagnet that show switching error in the presence of thermal noise. In some cases, a Skyrmion is formed (Fig. 4c, 10 ps) but subsequently moves towards the boundary (Fig. 4c, 17ps). Thus, the skyrmion is annihilated and ultimately returns to the ferromagnetic "up" state (Fig. 4c, 110ps), thereby failing to switch. The other type of error occurs when skyrmion cannot be formed at all. In this case, when PMA is reduced, the magnetization ends up in a multi-domain metastable state after the withdrawal of the voltage pulse (Fig. 4d, 110ps).

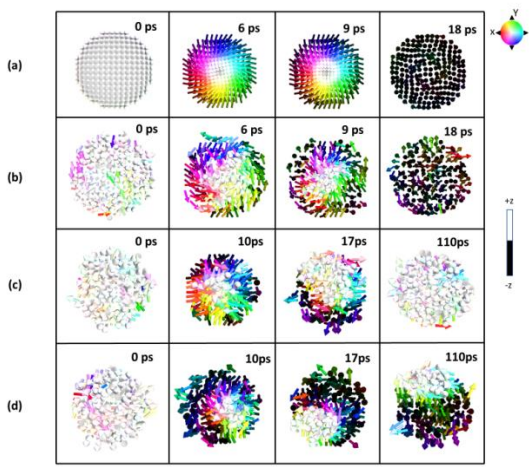

Fig. 4: Magnetization states visited at different time of 20 nm nanomagnets for (a) non-thermal switching (b) thermal switching, (c) and (d) error cases.

### C. Switching Energy Per Bit

Considering an application of a voltage pulse of 2.0 V for a 1.2 nm thick free layer and a 1 nm thick MgO layer with relative permittivity ~ 7, energy dissipated per switching event in a 100 nm nanomagnet is ~1fJ. Sub 1 fJ/bit switching for 20nm nanomagnet can be achieved (provided large VCMA co-efficient, PMA and DMI are available) which is highly desirable for energy efficient memory devices.

### IV. CONCLUSION

In summary, our simulations show that thermally robust switching with a frequency of 2 GHz can be attained for a 100 nm nanomagnet. This switching remains robust when scaling down to 50 nm lateral dimensions in the presence of thermal noise at ~20 GHz with appropriate choice of feasible material parameters based on published experimental studies [31,32,33].

While downscaling to 20 nm, switching with a very large frequency in the range of ~100 GHz can be achieved provided material parameters with large VCMA [34], DMI [35, 36] and PMA [34] are chosen. These material parameters (for the 20 nm nanomagnet simulation) have only been theoretically predicated [34, 35, 36] but not experimentally realized to date. Even with such parameters, the extremely small volume makes such nanomagnets prone to switching error in the presence of thermal perturbation.

Thus, skyrmion mediated ferromagnetic reversal offers an extremely energy efficient (with dissipation < 1 fJ/bit) alternative to spin transfer torque (STT) random access memory devices (with dissipation ~100 fJ/bit) while also offering competitive scaling to lateral dimensions ~50 nm and below. For aggressive scaling to 20 nm and below, ferrimagnet systems (rather than ferromagnets) could potentially offer a thermally robust skyrmion mediated switching route due to small stray fields and low DMI, PMA and VCMA requirement to form skyrmions [37,38].

### APPENDIX

### *I. Grid Size Dependence of Non-thermal Evolution of Topological Charge*

Switching dynamics depends on grid size. To observe the grid size dependence on switching dynamics, the 100nm MTJ is allowed to oscillate for 20 ns once the PMA has been reduced to 0.94 MJ/m$^3$ from 1.11MJ/m$^3$ within 0.1 ns. As the metastable skyrmion oscillates we compare the dynamics for 32×32×1, 64×64×1 and 100×100×1 grid. When the topological charge reaches the peak values PMA is restored to 1.11 MJ/m$^3$ and switching is found up to 9, 5 and 4 cycles for 32×32×1, 64×64×1 and

100×100×1 grid respectively (see Fig. 5). However, the dynamics are qualitatively similar for all cases. This shows that the "physics" of skyrmion reversal is affected by the boundary discretization due to different grid sizes and it is not a numerical issue with choice of grid size. In this study we choose 64×64×1 grid for 100 nm (each cell of 1.56nm×1.56nm×1.2nm) which closely mimics edge effects due to lithographic imperfections

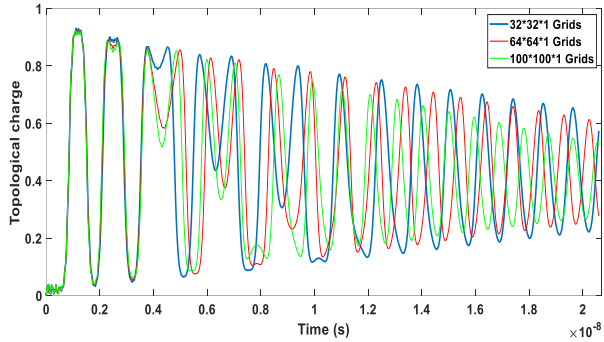

Fig. 5: Evolution of topological charge of 100nm MTJ for 32×32×1, 64×64×1 and 100×100×1 Grids

## II. Effect of Time Step Variation on Non-thermal Evolution of Topological Charge

To show the effect of varying time step on switching mechanism, non-thermal evolution of topological charge was observed. For 100nm MTJ and 64×64×1 grid the evolution of topological charge is shown in Fig. 6 where we can see for all time steps (100fs,50fs,10fs,5fs and adaptive), the time for formation of skyrmion is nearly same but the stability of metastable skyrmion differs. For 100fs time step, the metastable skyrmion gets destroyed as soon as the PMA is restored. For 50fs the skyrmion breathes for ~ 1ns after the PMA restoration. As the time step size is lowered the evolutionary pattern converges. Thus, we choose time step of 10 fs for all our simulations.

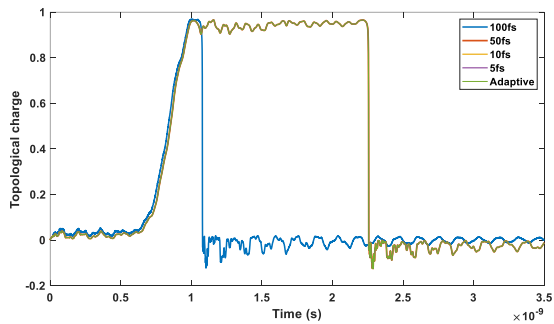

Fig. 6: Evolution of topological charge for different time steps